\documentclass[pdflatex,sn-mathphys-num]{sn-jnl}

\usepackage{graphicx}%
\usepackage{multirow}%
\usepackage{amsmath,amssymb,amsfonts}%
\usepackage{amsthm}%
\usepackage{mathrsfs}%
\usepackage[title]{appendix}%
\usepackage{xcolor}%
\usepackage{textcomp}%
\usepackage{manyfoot}%
\usepackage{booktabs}%
\usepackage{algorithm}%
\usepackage{algorithmicx}%
\usepackage{algpseudocode}%
\usepackage{listings}%
\usepackage{cite}                                             

\theoremstyle{thmstyleone}%
\theoremstyle{thmstyletwo}%

\theoremstyle{thmstylethree}%

\begin{document}

\title[Article Title]{Realization of a Chiral Photonic-Crystal Cavity with Broken Time-Reversal Symmetry}

\author[1,2]{\fnm{Kiran M.} \sur{Kulkarni}}
\equalcont{These authors contributed equally to this work.}
\author[3]{\fnm{Hongjing} \sur{Xu}}
\equalcont{These authors contributed equally to this work.}
\author[8]{\fnm{Fuyang} \sur{Tay}}
\author[1,2]{\fnm{Gustavo M.} \sur{Rodriguez-Barrios}}
\author[1,2]{\fnm{Dasom} \sur{Kim}}
\author[2,4]{\fnm{Alessandro} \sur{Alabastri}}
\author[5]{\fnm{Vasil} \sur{Rokaj}}
\author[6,7]{\fnm{Ceren B.} \sur{Dag}}
\author*[2,4,9]{\fnm{Andrey} \sur{Baydin}}\email{baydin@rice.edu}
\author*[2,3,4,9,10]{\fnm{Junichiro} \sur{Kono}}\email{kono@rice.edu}

\affil[1]{\orgdiv{Applied Physics Graduate Program, Smalley-Curl Institute}, \orgname{Rice University}, \orgaddress{\city{Houston}, \state{TX}, \country{USA}}}

\affil[2]{\orgdiv{Department of Electrical and Computer Engineering}, \orgname{Rice University}, \orgaddress{\city{Houston}, \state{TX}, \country{USA}}}

\affil[3]{\orgdiv{Department of Physics and Astronomy}, \orgname{Rice University}, \orgaddress{\city{Houston}, \state{TX}, \country{USA}}}

\affil[4]{\orgdiv{Smalley-Curl Institute}, \orgname{Rice University}, \orgaddress{\city{Houston}, \state{TX}, \country{USA}}}

\affil[5]{\orgdiv{Department of Physics}, \orgname{Villanova University}, \orgaddress{\city{Villanova}, \state{PA} \country{USA}}}

\affil[6]{\orgdiv{Department of Physics}, \orgname{Indiana University}, \orgaddress{\city{Bloomington}, \state{IN}, \country{USA}}}

\affil[7]{\orgdiv{Department of Physics}, \orgname{Harvard University}, \orgaddress{\city{Cambridge}, \state{MA}, \country{USA}}}

\affil[8]{\orgdiv{Department of Physics}, \orgname{Columbia University}, \orgaddress{\city{New York}, \state{NY}, \country{USA}}}

\affil[9]{\orgdiv{Rice Advanced Materials Institute}, \orgname{Rice University}, \orgaddress{\city{Houston}, \state{TX}, \country{USA}}}

\affil[10]{\orgdiv{Department of Materials Science and NanoEngineering}, \orgname{Rice University}, \orgaddress{\city{Houston}, \state{TX}, \country{USA}}}

\abstract{
Light--matter interactions in chiral cavities offer a compelling route to manipulate material properties by breaking fundamental symmetries such as time-reversal symmetry. However, only a limited number of chiral cavity implementations exhibiting broken time-reversal symmetry have been demonstrated to date. These typically rely on either the application of strong magnetic fields, circularly polarized Floquet driving, or the hybridization of cavity modes with matter excitations in the ultrastrong coupling regime. Here, we present a one-dimensional terahertz photonic-crystal cavity that exhibits broken time-reversal symmetry. The cavity consists of a high-resistivity silicon wafer sandwiched between lightly \textit{n}-doped InSb wafers. By exploiting the nonreciprocal response of a terahertz magnetoplasma and the exceptionally low effective mass of electrons in InSb, we demonstrate a circularly polarized cavity mode at 0.67\,THz under a modest magnetic field of 0.3\,T, with a quality factor exceeding 50. Temperature-, magnetic field-, and polarization-dependent measurements, supported by simulations, confirm the realization of a chiral cavity with broken time-reversal symmetry. This platform offers a robust and accessible approach for exploring chiral light--matter interactions and vacuum dressed quantum condensed matter in the terahertz regime.
}

\maketitle

\section*{Introduction}\label{intro}
Recent advancements in materials science and photonics have opened myriad research opportunities in the study of light--matter interactions in condensed matter systems. Strong coupling between electromagnetic cavity modes and various materials has emerged as a powerful platform for generating new quantum phases and nontrivial ground states under equilibrium conditions by dressing a material’s excitations with the enhanced vacuum fluctuations of the cavity field~\citep{baydin2025quantum, peraca2020ultrastrong, forn2019ultrastrong, frisk2019ultrastrong}. Recent theoretical developments have demonstrated that harnessing symmetry breaking in these systems unlocks access to exotic material properties~\citep{forn2010observation, baumann2010dicke, garziano2014vacuum}. Symmetries not only govern conservation laws—they also play a fundamental role in shaping a material's emergent behavior~\citep{wehling2014dirac, lu2014topological, kasprzak2006bose}. Preserving or explicitly breaking a symmetry provides additional degrees of freedom for controlling light--matter interactions. For example, breaking translational symmetry—by introducing defects in photonic crystals~\citep{tanese2013polariton, jacqmin2014direct}, twisting van der Waals layers~\citep{andrei2020graphene}, or engineering artificial superlattices~\citep{regan2022emerging, xin2022twist, regan2020mott}—
can form novel collective excitations and reshape the underlying band structure (photonic, electronic, or hybrid), thereby opening new avenues for enhanced light–matter coupling.

Traditionally, strong driving fields have been employed to break symmetries—such as Floquet engineering~\citep{oka2019floquet, rudner2019floquet}—that periodically perturb materials to reveal hidden phases and ground states not accessible in equilibrium. More recently, a complementary approach has emerged that leverages the quantum properties of electromagnetic modes in optical cavities~\citep{sentef2020quantum}. In the cavity quantum electrodynamics framework, a material’s electronic states can be dressed by cavity vacuum fluctuations rather than by external laser fields. In this setting, especially in the ultrastrong coupling (USC) regime, cavity-enhanced vacuum fluctuations can significantly modify collective excitations, electron--phonon interactions, and even the ground state of the material without requiring any external drive~\citep{schlawin2022cavity, garcia2021manipulating}. Recent studies suggest that by engineering the symmetry properties of these vacuum fields, one can effectively imprint symmetry-broken states onto the material ~\citep{appugliese2022breakdown, ashida2020quantum, ke2023vacuum}.

There have been significant efforts in developing cavities that support circularly polarized modes by breaking time-reversal symmetry (TRS) via magneto-optic (Faraday) bias or spatiotemporal modulation~\citep{caloz2018electromagnetic}, while simultaneously ensuring ultrastrong light-matter coupling. Nonreciprocal rings and isolators relying on clockwise/counter-clockwise splitting waves and multi-pass rotation have been demonstrated; however, they do not realize a confined and uniformly circular chiral mode~\citep{yuan_-chip_2021, ji_terahertz_2020}. 
In this context, cavity implementations relevant to this objective should confine a spatially uniform circularly polarized cavity mode, break TRS while maintaining high \(Q\)-factors and small effective mode volume, and, importantly, the chirality should arise from the photon mode itself rather than from light-matter hybridization. 

A spatially uniform circularly polarized cavity mode means near-unity ellipticity of the cavity mode throughout the mode confinement region, not merely on average. Fabry-Pérot and distributed Bragg reflector designs has partially achieved this goal by using Faraday rotation~\citep{silverman1994interferometric}, polarization-dependent Fano responses~\citep{piao2015spectral}, metasurfaces that set different loss rates for the two helicities~\citep{overvig_chiral_2021}, and photonic-crystal mirrors that provide large chiral optical response (mitigating the typical inversion caused by standard mirror interfaces)~\citep{plum2015chiral, voronin2022single, semnani_spin-preserving_2020}. Among these, Suárez-Forero et al. realize an optical cavity with magnetically biased atomically thin mirrors that select helicity; the cavity mode is laterally extended and lies in the optical band~\citep{suarez2024chiral}. These approaches control helicity, but the mode volume is large, operation is often outside the THz range, or a high magnetic field is required. Additionally, chiral BIC (bound states in continuum) and quasi-BIC metasurfaces achieve high \(Q\)-factors by suppressing radiation and show strong response to circularly polarized light, but the resonant mode spans the metasurface area--limiting effective mode volume--and spatially non-uniform ellipticity~\citep{overvig_chiral_2021, lv_robust_2024}. In the THz range, Landau polaritons in sub-wavelength resonators over high-mobility 2DEGs produce strong dichroism but yield modest \(Q\)-factors, with the chirality tied to the hybridized polariton mode~\citep{andberger2024terahertz}, while surface-plasmon-based patch cavities on gyrotropic substrates provide strong confinement with the helicity governed by the magnetoplasmonic near field rather than a uniform circular photonic mode~\citep{aupiais2024chiral}. 

Here, we demonstrate a chiral photonic-crystal cavity (PCC) with broken TRS, achieved by replacing isotropic layers in a Bragg stack with a gyrotropic material—$n$-doped InSb—under an external magnetic field of 0.3\,T, as recently predicted in our prior theoretical work~\citep{tay2025terahertz}. In contrast to the above-mentioned demonstration, our chiral cavity supports a uniform circular polarized cavity mode in the THz range and features the mode volume and high Q-factor that are similar to conventional PCCs, which were previously used to achieve the USC regime. Thus, it becomes a promising platform for chiral vacuum light--matter interaction research. 

Our cavity design selectively suppresses one circular polarization, corresponding to the cyclotron-resonance-active (CRA) mode that co-rotates with the electron cyclotron motion. However, the opposite cyclotron-resonance-inactive (CRI) mode that counter-rotates with the electron is preserved with minimal losses. In this proof-of-concept implementation, we achieved a $Q$-factor exceeding 50, with potential for further enhancement. We characterized the cavity using terahertz time-domain spectroscopy (THz-TDS), which confirmed the predicted polarization-selected response. The cavity operates robustly over a broad temperature ($T$) range and enables tunability through the application of a magnetic field ($B$), albeit within a defined range. Our platform introduces new functionality to a standard cavity architecture commonly used in USC studies, enabling chiral photonic modes with broken TRS. This opens promising avenues for investigating topological effects, nonreciprocal waveguiding, and emergent quantum phases mediated by cavity-engineered quantum vacuum fluctuations~\citep{tay2025terahertz, khanikaev2013photonic, lodahl2017chiral, bliokh2015quantum, karzig2015topological}.

\section*{Results}
\subsection*{Designing a chiral cavity}
Gyrotropic materials exhibit magneto-optical effects, i.e., nonreciprocal light propagation, due to their optical properties. For example, the motion of free carriers in a plasma in $B$ leads to CR and, consequently, circular dichroism. 
The off-diagonal element of the gyrotropic permittivity introduces anisotropy in the magnetoplasma, making the permittivity tensor asymmetric~\citep{PalikFurdyna1970RPPa}. This asymmetry, which exists as a result of broken TRS due to the applied $B$, differentiates the propagation of the CRA and CRI modes. As proposed in our earlier theoretical  work~\citep{tay2025terahertz}, we chose to focus on lightly doped InSb, a narrow band gap semiconductor with a small electron effective mass. 
\begin{figure}
    \centering
    \includegraphics[width=1\linewidth]{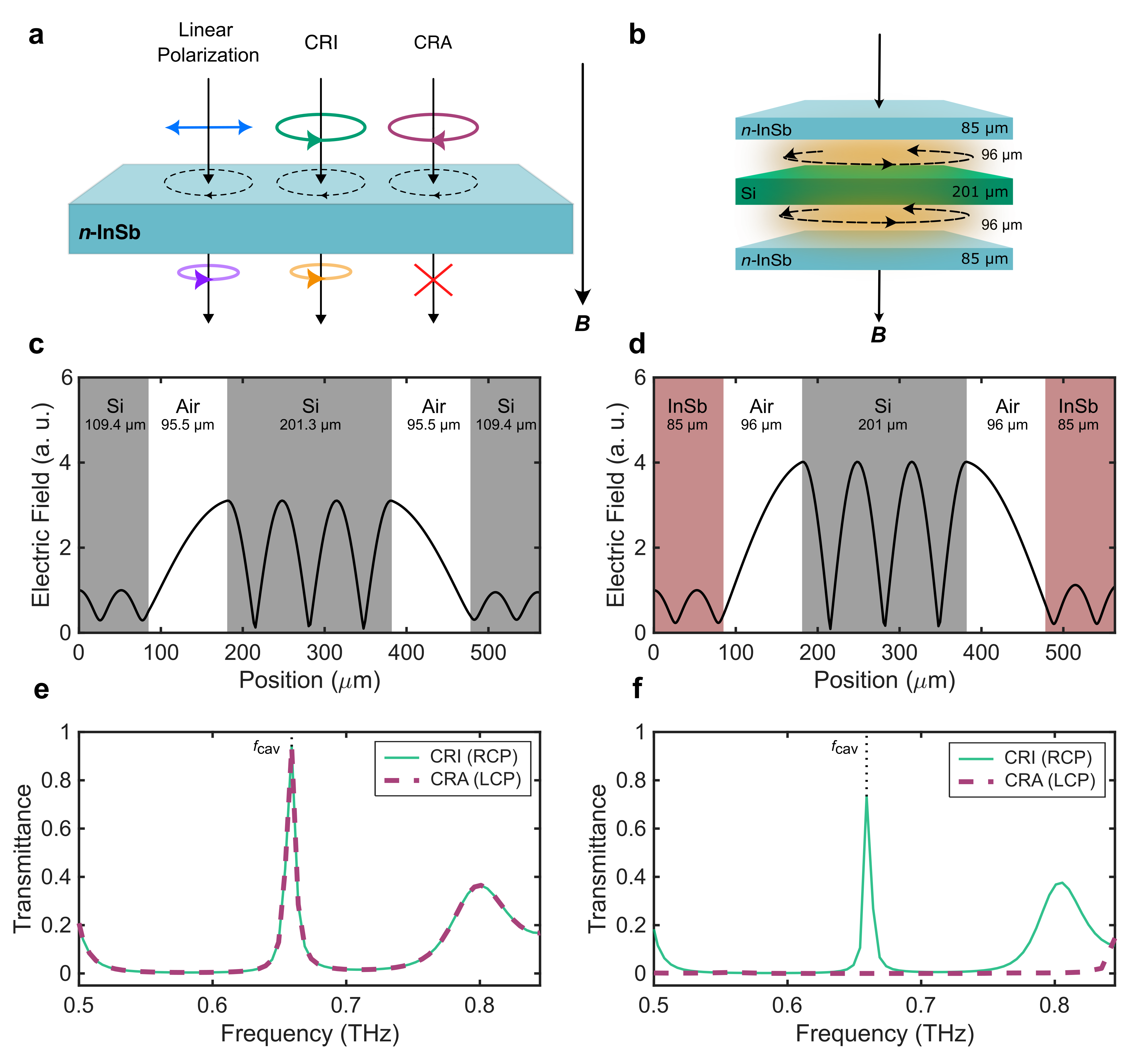}
    \caption{\textbf{Chiral Photonic Crystal Cavity Design.} \textbf{a},~Schematic of light interaction with a chiral material: in the presence of a magnetic field, the material selectively transmits the CRI mode while absorbing the CRA mode. \textbf{b},~Schematic of the designed chiral PCC: the PCC is composed of alternating InSb and air layers with a central silicon defect layer. The external magnetic field applied normal to the InSb layers imparts chiral, nonreciprocal behavior. \textbf{c},~Simulated electric field distribution for the conventional Si-based 1D-PCC: normalized electric field intensity as a function of position in a nonchiral PCC (with only Si layers). Standing-wave patterns arise from Bragg reflections within the periodic layers. The defect layer (the middle layer) introduces high field-confinement, characteristic of resonant modes confined within the photonic band gap. \textbf{d},~Simulated electric field distribution for the chiral PCC. \textbf{e},~ Simulated transmission spectrum for the conventional Si-based 1D-PCC: transmittance versus frequency reveals identical responses for the RCP (= the CRI mode) and LCP (= CRA) modes, reflecting the isotropic, non-chiral nature of Si layers. \textbf{f},~Simulated transmission spectrum for the chiral 1D-PCC: a pronounced resonance emerges near 0.66\,THz for CRI, while the CRA transmission is strongly suppressed. This polarization-selective behavior validates the chiral nature of the cavity design. Here, the cavity frequency is depicted by the dotted line at 0.66\,THz.}
    \label{fig:Chiral_PCC}
\end{figure}

Figure~\ref{fig:Chiral_PCC}a illustrates the nonreciprocal transmission of the CRA and CRI modes through InSb in a $B$ applied perpendicular to the surface. For \(B > 0\), the left-handed circularly polarized (LCP) mode corresponds to CRA, and the right-handed circularly polarized (RCP) mode to CRI. Namely, InSb can selectively transmit a circularly polarized mode based on the direction of the magnetic field. Thus, by sandwiching a Si wafer between two InSb wafers as depicted in Figure~\ref{fig:Chiral_PCC}b, we can create a chiral PCC, where only one circular polarization is confined. 

In conventional Si-based PCC designs in the THz range that utilize Si wafers separated by air gaps (Figure~\ref{fig:Chiral_PCC}c), both the LCP and RCP light experience the same refractive index, irrespective of the magnetic field direction, and propagate identically, exhibiting reciprocal behavior. When Si wafers are all the same thickness, it will result in a photonic band gap (or a stopband) due to the Bragg reflections within the alternating Si-air layers. The introduction of a defect layer (the thickness of the center Si wafer is larger than the Si side wafers) breaks the periodicity of the stack and creates a localized mode, confining standing waves at specific frequencies in the stacking direction of the wafers, which is located inside the photonic band gap. Figure~\ref{fig:Chiral_PCC}e depicts the cavity mode at \(f_\textrm{cav} = 0.66\)\,THz for the Si-based PCC shown in Figure~\ref{fig:Chiral_PCC}c. The simulated electric field distribution corresponding to this mode is plotted in Figure~\ref{fig:Chiral_PCC}c, which is maximized at the surface of the center Si wafer. 

Replacement of the periodic layers in the Si-based 1D-PCC with gyrotropic materials such as InSb alters the refractive index profile for each circular polarization mode~\citep{lee2014defect, aly2017tunable,tay2025terahertz, arikawa2012giant, ju2023creating}. We designed a chiral 1D-PCC by substituting the side Si wafers with lightly $n$-doped InSb wafers. Figure~\ref{fig:Chiral_PCC}b shows our designed chiral PCC structure with experimental wafer thicknesses. We keep the center Si wafer unchanged, as it will be a substrate for thin film samples in future studies. Each layer thickness is chosen to satisfy the Bragg interference condition, with an optical thickness corresponding to one or more quarter wavelengths at the design frequency (Figure \ref{fig:Chiral_PCC}d), producing a photonic band gap between 0.5\,THz and 0.8\,THz, as shown in the simulated transmission spectrum in Figure~\ref{fig:Chiral_PCC}f. 

The complex permittivity of InSb for the CRA \((+)\) and CRI \((-)\) modes is given by: \(\tilde{\epsilon}_\pm =\epsilon_{xx} \pm i \epsilon_{xy}\)~\citep{tay2025terahertz}. For \(B>0\), the imaginary part of the permittivity for the CRA mode (LCP) exhibits a Lorentzian peak centered at the cyclotron frequency \(f_\text{c} = \omega_\text{c}/2\pi\). Thus, the CRA mode near \(f_\text{c}\) is strongly absorbed, while the CRI (RCP) mode experiences negligible absorption because Im\(\{\tilde{\epsilon}_{-}\} \approx 0\) in the same frequency range. Moreover, the real part of \(\tilde{\epsilon}_{-}\) for CRI (RCP) approximates the permittivity of Si, \(\tilde{\epsilon}_\mathrm{Si}\). Replacement of Si layers with InSb layers (Figure~\ref{fig:Chiral_PCC}d) therefore, yields a cavity that retains similar photonic band gap properties for the CRI (RCP) mode. A simulated transmission spectrum for the designed chiral PCC is shown in Figure~\ref{fig:Chiral_PCC}f. Due to the strong absorption of the CRA (LCP) mode near the cyclotron frequency, the cavity becomes chiral: the cavity mode vanishes for the CRA (LCP) mode but persists for the CRI (RCP) mode.

\subsection*{Experimental Characterization of the Cavity}
To validate the simulations and assess the practical performance of the chiral PCC, we conducted terahertz time-domain spectroscopy (THz-TDS) measurements at various magnetic fields and temperatures in the Faraday geometry. The cavity was fabricated by first polishing InSb and Si wafers to precise thicknesses and then stacking them with designed spacings. For more details, see the Methods section.

\begin{figure}[htb]
    \centering
    \includegraphics[width=0.9\linewidth]{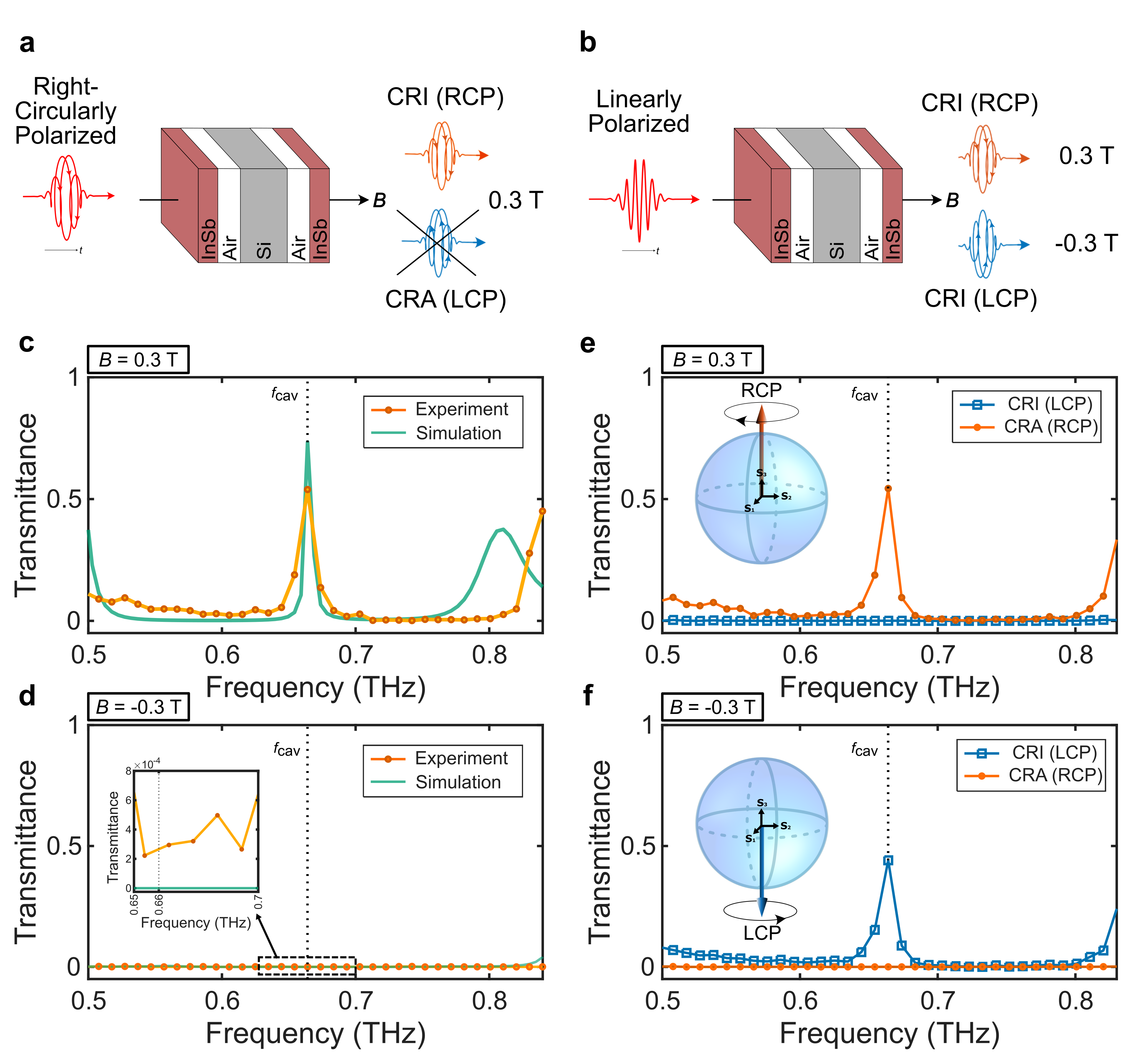}
    \caption{\textbf{Experimental Demonstration of a THz Chiral Photonic-Crystal Cavity.} \textbf{a} and \textbf{b},~interaction of circularly and linearly polarized THz radiation, respectively, with the chiral cavity at $B=\pm0.3\,\mathrm{T}$. The chiral PCC selectively absorbs the CRA (LCP) mode while transmitting the CRI (RCP)  mode (for \(B > 0\)). Linearly polarized radiation consists of 50\% RCP and 50\% LCP components. The RCP (CRI) component is transmitted for $B=0.3\,\mathrm{T}$, while the LCP (CRA) component is transmitted for $B = -0.3\,\mathrm{T}$, validating the designed selective circular dichroism. \textbf{c} and \textbf{d},~Simulated (jade) and experimental (orange) spectra for RCP (CRI) at $B=\pm0.3\,\mathrm{T}$ and $T = 4\,\mathrm{K}$. The transmission peak is observed at 0.66\,THz, highlighted by a dotted vertical line, for $B=0.3\,\mathrm{T}$ is the cavity mode. This mode is absent when $B = -0.3\,\mathrm{T}$, reflecting the suppression of the incident RCP (CRA) mode. \textbf{e} and \textbf{f},~Experimental spectra for linearly polarized THz radiation incidence for $B=\pm0.3\,\mathrm{T}$ at $T = 4\,\mathrm{K}$. 
    At $B=0.3\,\mathrm{T}$ ($-0.3\,\mathrm{T}$), a distinct resonance peak is observed near 0.66\,THz for the RCP (LCP) mode, while the LCP (RCP) mode shows no significant transmission at this frequency. Insets show the normalized Stokes vector \(S/S_0\) on the Poincaré sphere (using Eq. \ref{Eq: Stokes}), with the arrow marking the cavity’s polarization state; the north (south) pole corresponds to RCP (LCP) modes. Here, the cavity frequency is depicted by the dotted line at 0.66\,THz.
    }
    \label{fig:exp}
\end{figure}

Two experimental configurations were used to study the chiral PCC, as illustrated in Figure~\ref{fig:exp}a and Figure~\ref{fig:exp}b.
First, we characterized the response of the chiral cavity to RCP THz radiation, which was generated by passing a linearly polarized THz beam through a quarter-wave plate. Figure~\ref{fig:exp}a shows schematically what we expect when RCP radiation traverses the chiral cavity. In a positive magnetic field, the transmittance should not be affected, but in a negative magnetic field, the transmittance should be suppressed. Figures~\ref{fig:exp}c and \ref{fig:exp}d display measured transmission spectra for an RCP beam for \(B = +0.3\)\,T and \(-0.3\)\,T, respectively, at a temperature of \(4\)\,K. It can be seen that the cavity defect mode at \(0.66\)\,THz persists for a positive magnetic field (Figure~\ref{fig:exp}c) while it is suppressed for a negative field (Figure~\ref{fig:exp}d). This behavior confirms the chiral cavity's selectivity of transmitting RCP radiation only in a positive magnetic field, i.e., a field applied along the propagation direction.

Next, we tested the chiral cavity with linearly polarized THz radiation, which contains 50\% RCP and 50\% LCP. In this case, we expect only the RCP (LCP) component to be transmitted when the applied magnetic field is positive (negative), as depicted in Figure~\ref{fig:exp}b. Figures~\ref{fig:exp}e and \ref{fig:exp}f display experimental transmittance spectra for a linearly polarized THz beam input. It can be seen that the RCP (LCP) peak is retained under the application of a positive (negative) magnetic field. Again, this observation verifies the circular polarization selectivity of the cavity. The inset in each figure shows that the experimentally obtained ellipticity angle approaches the theoretical limit of \(\pm\pi/4\). 

More experimental data at $T=4\,\mathrm{K}$, $70\,\mathrm{K}$, and $100\,\mathrm{K}$ are shown in the Supplementary Material. 

\subsection*{Temperature and magnetic field dependence}

\begin{figure}
    \centering
    \includegraphics[width=1\linewidth]{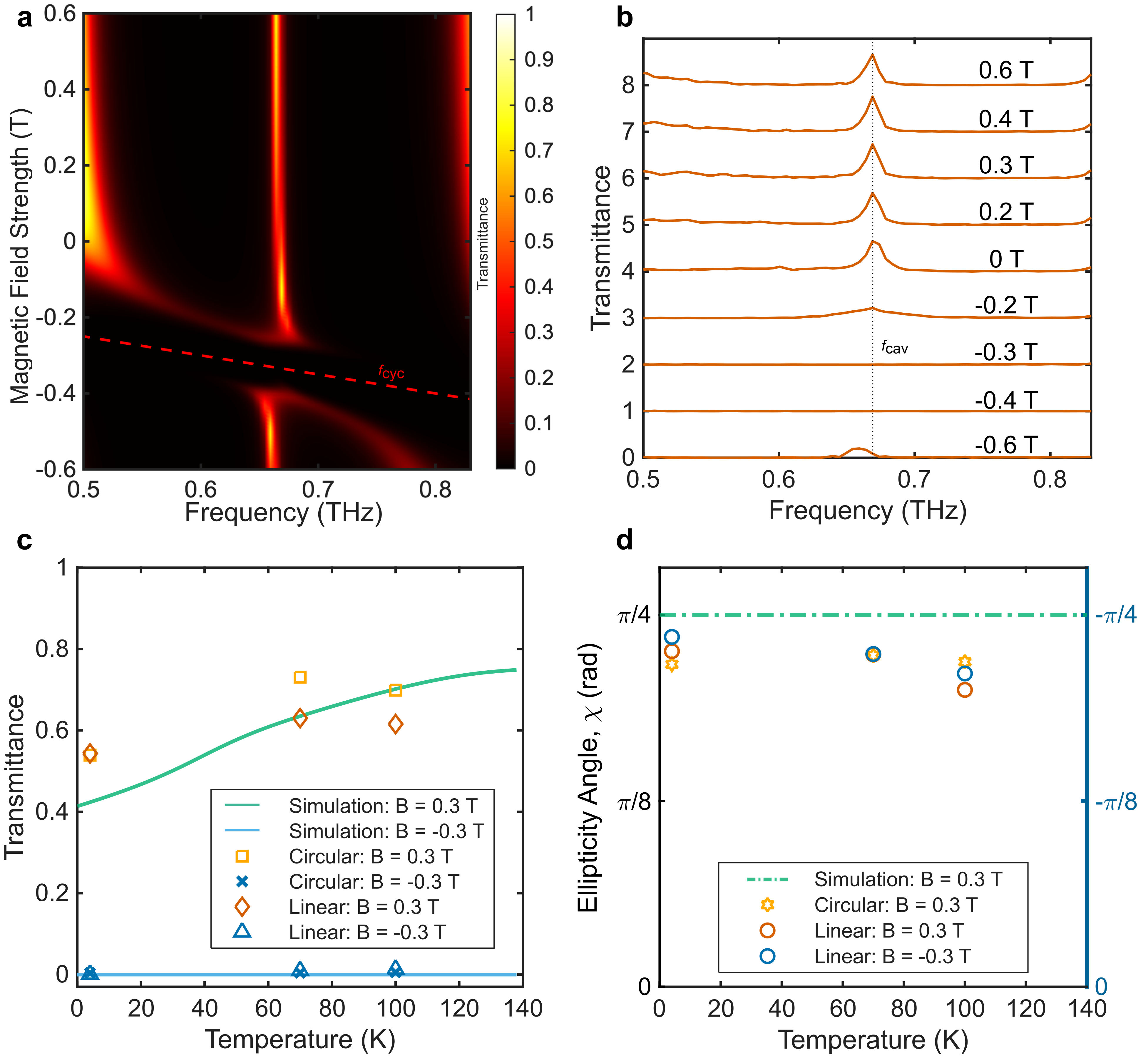}
    \caption{\textbf{Tunability of the Chiral PCC.} \textbf{a},~Simulated 2D color map depicting the transmittance versus frequency and magnetic field at $T = 70\,\mathrm{K}$ for incident RCP radiation. A prominent cavity mode is observed at 0.66\,THz for positive magnetic fields, while a stopband is observed for \(B = -0.2\)\,T to \(-0.4\)\,T, showcasing the operational range and sign of \(B\) for the chiral PCC. Here, the cyclotron frequency of the magnetoplasma is depicted by the red dashed line. \textbf{b},~Experimental transmittance spectra for RCP at $T = 70\,\mathrm{K}$: Experimental transmission spectra verify the simulation, showcasing suppression of the RCP mode for \(B = -0.2\)\,T to \(-0.4\)\,T. Here, the cavity frequency is depicted by the dotted line at 0.66\,THz. \textbf{c},~Temperature-dependent transmittance spectra showing the cavity mode at 0.66\,THz, indicating robust mode confinement over a vast range of temperature. \textbf{d},~Temperature-dependent ellipticity angle, \(\chi\). It converges to \(\pi/4\) for both circular and linear polarized beams at \(B = 0.3\)\,T while, for \(B = -0.3\)\,T, it converges to \(-\pi/4\).}
    \label{fig:tunability}
\end{figure} 

We carried out $B$-dependent simulations and measurements. Figure~\ref{fig:tunability}a shows a RCP transmittance colormap for the chiral cavity as a function of magnetic field and frequency at \(T = 70\)\,K. The CR shifts with $B$, as evidenced by the fully-gapped stopband in the spectrum appearing when $-0.4\,\mathrm{T} < B < -0.2\,\mathrm{T}$ while the cyclotron frequency $f_\text{c}$ passes through the stopband frequency range. Notably, the cavity effectively suppresses the CRA mode within the range \(-0.4\,\text{T} < B< -0.2\,\text{T}\). This behavior is confirmed by experimental RCP transmission spectra in Figure~\ref{fig:tunability}b, which show that the cavity mode at \(f_\text{cav} \approx 0.66\)\,THz is suppressed for \(-0.4\,\text{T} < B< -0.2\,\text{T}\). These findings establish the optimal magnetic field range for the effective operation of the chiral cavity. The usable range is set by the width of the CRA stop band (see Figure \ref{fig:tunability}a), which follows the Lorentzian profile of \(\rm{Im}\{\rm\epsilon_{\rm{CRA}}\}\) (see Supplementary Material). Thus, even a lower $B$ (than the design) can be used to suppress one circular polarization. 

We characterized the performance of a chiral cavity that was optimally designed to work for 70\,K based on $T$-dependent InSb's refractive index~\citep{ju2021tunable} over a range of $T$. At room temperature, the chosen InSb wafers become highly opaque due to an increased intrinsic free-carrier density being thermally excited across the band gap, which significantly reduces transmission through the cavity~\citep{ulbricht2011carrier}. As $T$ decreases, fewer carriers are thermally excited, which reduces absorption. However, at very low $T$, scattering due to impurities increases the scattering rate and reduces transmission.
These two effects—thermal free-carrier absorption at high $T$ and impurity scattering at low $T$—compete, resulting in an optimum $T$ where transmission is maximized and scattering is minimized. In our experiments, this optimal $T$ was found to be around \(T_{\text{opt}} = 70\,\text{K}\). The free-space characterization of the InSb and the complex permittivity and refractive index is reported in the Supplementary Material. 

Figure~\ref{fig:tunability}c shows the transmittance of the cavity mode as a function of $T$. One can see that the transmittance does not change significantly in the range between 0 and 150\,K. Simulations and experiments are also in agreement. Above 150\,K, the transmittance drops significantly due to absorption by intrinsic free carriers thermally excited across the band gap. The ellipticity value, on the other hand, is constant with $T$, as seen in Figure~\ref{fig:tunability}d. The experimental ellipticity angles measured at 4\,K, 70\,K, and 100\,K are near the theoretical maximum of \(\pm\pi/4\), confirming high circular dichroism in a wide temperature range below or near the optimal temperature of \(T_{\text{opt}} = 70\,\text{K}\). The experimental data and calculation details of the Stokes parameter and ellipticity are provided in the Methods section and the Supplementary Material. 

\section*{Discussion and Outlook}\label{conclusions}
From a practical perspective, exploiting symmetry breaking in the cavity-QED systems enables the design of next-generation quantum technologies, including robust quantum simulators, non-reciprocal light sources, and ultrasensitive quantum sensors \citep{koch2010time, yang2025emergent, metelmann2015nonreciprocal, lodahl2017chiral, roushan2017chiral, maleki2023time, hauff2022chiral}. Our chiral cavity exhibits a promising avenue to aid in these developments. The experimental demonstration of the chiral PCC shows good agreement with the theoretical simulations performed using the Maxwell equations. The transmission spectra obtained for circularly polarized modes and their dependence on the magnetic field strength and direction align with the predicted behavior of the chiral PCC. While the $Q$-factor of the cavity is already high, it can be further enhanced by using multiple periods of InSb and air layers on either side of the central defect layer. Additionally, the high efficiency of our chiral cavity in the broad temperature range underscores its potential as a robust platform for probing chiral light-matter interactions in condensed matter systems.

Beyond verifying the fundamental properties of our design, these results point to broader avenues for manipulating time-reversal symmetry in equilibrium—akin to Floquet-engineered phases but without requiring a strong external drive. For instance, chiral-light irradiation in graphene can open an anomalous Hall gap, and embedding graphene in a chiral cavity may induce topological phases in the ground state through virtual photon exchange~\citep{kibis2011band, dag2024engineering, wang2019cavity, appugliese2022breakdown, yang2025quantum, jiang2024engineering, ghorashi_tunable_2025}. Likewise, Dirac/Weyl materials could be altered by chiral-cavity-induced phase transitions, while Berry curvature and nonlinear susceptibilities might be tuned via cavity-enhanced interactions~\citep{masuki2023berry, basov2017towards, schlawin2022cavity, jiang2025angular, yang2025emergent}. Moreover, in the USC regime, one anticipates ground-state modifications due to intrinsic squeezing~\citep{schwendimann1992nonclassical, ciuti2005quantum}, offering new opportunities for topological and nonreciprocal photonic devices. 
Future research can thus explore integrating active or low-dimensional materials within this chiral cavity for tunable polarization control and probing these emergent phenomena at the intersection of chiral quantum optics, topological transport, and strong light--matter coupling.

\section*{Methods} \label{Section: Methods}
\subsubsection*{Cavity fabrication}
The chiral PCC was assembled by stacking polished InSb and Si wafers with thicknesses determined by the Bragg condition and quarter-wave stack requirement. Air gaps between the InSb layers were maintained using paper spacers to achieve the desired optical path lengths. The thickness of each layer was verified with a micrometer.

\subsubsection*{THz time-domain magnetospectroscopy}
Time-domain THz magneto-spectroscopy measurements were performed in the Faraday geometry, where the direction of the applied magnetic field was parallel to the THz beam and the stacking direction of the cavity. We used a Spectra-Physics Solstice ultrafast amplifier (800\,nm wavelength, 1\,kHz repetition rate, 35\,fs pulse duration, 7\,W average power) to generate linearly polarized THz pulses (0.25\,THz\textendash2.5\,THz) via optical rectification in ZnTe. The THz beam was then sent through the center of an Oxford Instruments Spectromag superconducting magnet (1.4\textendash300\,K, 0\textendash10\,T), and transmitted through the sample. It was then detected in another ZnTe using electro-optic sampling. For each temperature and magnetic field setting, we collected the time-domain signal for delays up to 200\,ps with a time step as small as 200\,fs. The parameters were chosen so that as many echoes as possible were recorded before becoming indistinguishable from noise, and the time step was short enough to sample the highest frequency (2.5\,THz) within the Nyquist limit.

\subsubsection*{Determination of the polarization state of light}
Two experimental configurations were employed to study the chiral PCC, as illustrated in Figure~\ref{fig:exp}a and Figure~\ref{fig:exp}b. A linearly polarized THz beam passes through a THz quarter-wave plate (QWP) to generate RCP THz radiation before interacting with the chiral PCC. The transmitted light is analyzed using linear polarizers oriented at \(+45^\circ\) and \(-45^\circ\). From these components, the $x$- and $y$-components of transmitted radiation, $E_{x}$ and $E_{y}$, are calculated~\citep{rodriguez-barrios_terahertz_2025}. In another experiment, the QWP is removed, and a linearly polarized THz beam directly illuminates the chiral PCC. The transmitted light is analyzed with polarizers at \(+45^\circ\) and \(-45^\circ\) as before. Ellipticity calculations based on these measurements assess the cavity's ability to convert linearly polarized light into circularly polarized light depending on the direction of the magnetic field.

The ellipticity angle (\(\chi\)), is calculated from the Stokes parameters~\citep{stokes1851composition} as
\begin{align}
    \label{Eq: Stokes}
    S_0 &= {|E_x|^2 + |E_y|^2} \nonumber \\
    S_1 &= {|E_x|^2 - |E_y|^2} \nonumber\\
    S_2 &= {E_x E_y^* + E_y E_x^*} \nonumber\\
    S_3 &= {i \left(E_x E_y^* - E_y E_x^*\right)} 
\end{align}
\begin{equation}
    \label{Eq: Eta}
    \chi  = -\frac{1}{2}\arcsin{\frac{S_3}{S_0}} = -\frac{1}{2}\arcsin{\frac{i(E_x E_y^* - E_y E_x^*) }{|E_x|^2 + |E_y|^2}}
\end{equation}

More details can be found in the Supplementary Material. 

\backmatter


\bmhead{Acknowledgements}

We acknowledge Ikufumi Katayama for lending us the QWP. 

\section*{Declarations}

\begin{itemize}
\item Funding
    This research was developed with funding from the Defense Advanced Research Projects Agency (DARPA) Advanced Research Concepts (ARC) QUAMELEON Opportunity. The views, opinions, and/or findings expressed are those of the author(s) and should not be interpreted as representing the official views or policies of the Department of Defense or the U.S. Government. J.K. also acknowledges the U.S. Army Research Office (through Award No. W911NF2110157), the Gordon and Betty Moore Foundation (through Grant No. 11520), the W. M. Keck Foundation (through Award No. 995764), and the Robert A. Welch Foundation (through Grant No. C-1509). A.B., V.R., C.D.: This research was supported in part by grant NSF PHY-2309135 to the Kavli Institute for Theoretical Physics (KITP). 

\item Conflict of interest/Competing interests

    The authors declare no competing interests.

\item Ethics approval and consent to participate

    Not applicable.

\item Consent for publication

    Not applicable.

\item Data availability 

    The data that support the plots in this paper are available from the corresponding author upon reasonable request.


\item Code availability 

    The codes used in this study are available from the corresponding author upon reasonable request.

\item Author contribution
    F.T., A.B., and J.K. conceptualized the project. K.M.K. designed the cavity and performed numerical simulations with input from A.B. K.M.K., and A.B. fabricated the cavity. H.X. characterized the cavity with input from K.M.K, G.R.B., and A.B. D.K. built the setup. G.R.B. optimized the setup for ellipticity measurements. A.B. and J.K. supervised the project. K.M.K., H.X., A.B., and J.K. prepared the manuscript with inputs from all authors.
\end{itemize}

\noindent

\bibliography{ 2-references}
\renewcommand{\thefigure}{S\arabic{figure}} 




\raggedbottom

\setcounter{figure}{0}
\section*{\Large Supplementary Material}

In this supplementary material, we provide additional detail on the characterization of InSb wafers, the chiral cavity at various temperatures, and ellipticity calculations.  




\section*{InSb characterization}
\begin{figure}
    \centering
    \includegraphics[width=1\linewidth]{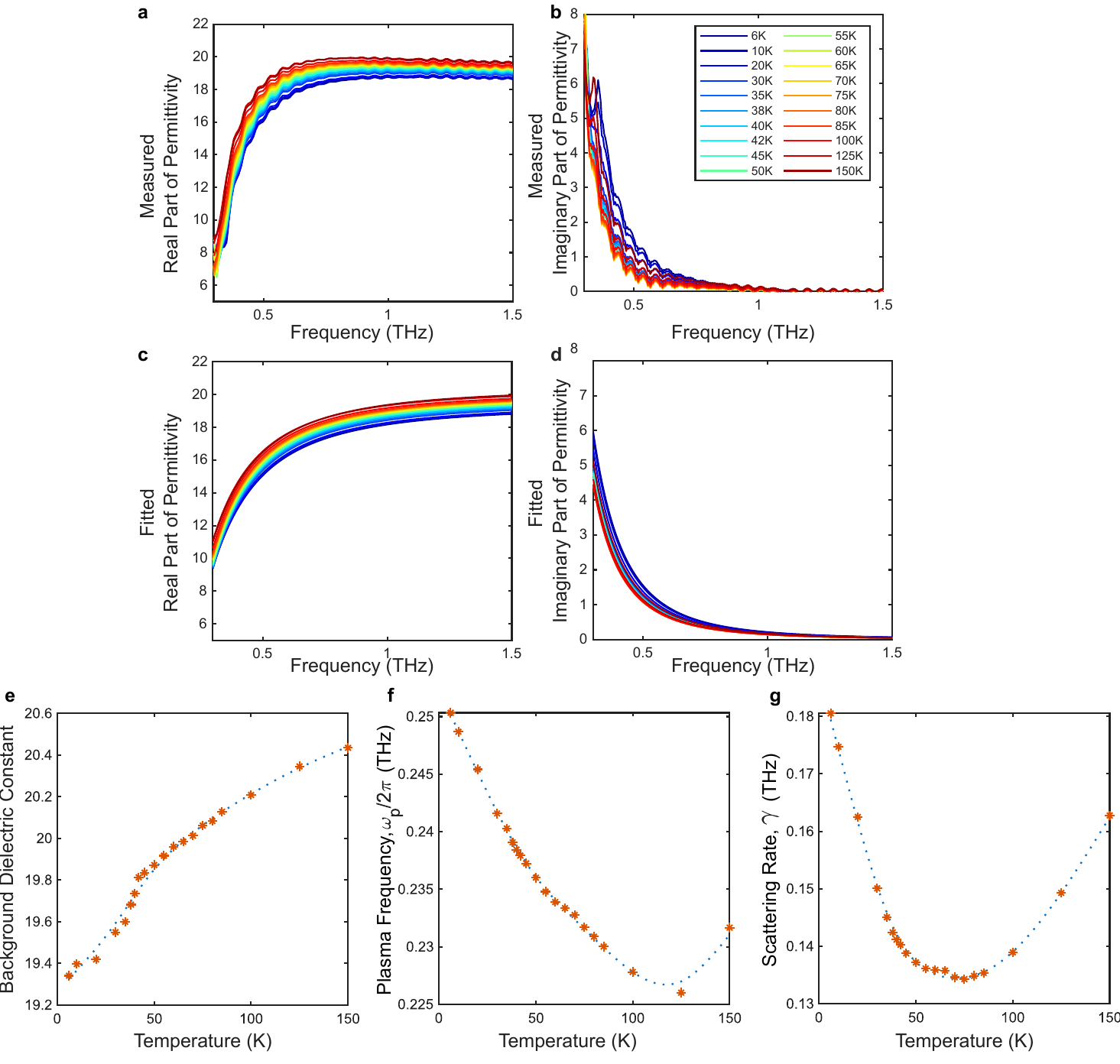}
    \caption{\textbf{Temperature-dependent complex permittivity of InSb in free space.} \textbf{a and b},~The real and imaginary part of the complex permittivity of InSb in free space at $B=0\,$T and various temperatures, obtained experimentally by THz-TDS. \textbf{c and d},~ Fitted Drude complex permittivity from the experiment at $B=0\,$T. \textbf{e-g},~ The fitted Drude parameters and their temperature dependence (the dotted line is the guide for the eye).}
    \label{fig:InSb_temperature}
\end{figure}

We characterized the temperature-dependent complex permittivity of InSb by terahertz time-domain spectroscopy (THz-TDS) in free space at $B=0\,$T, as shown in Fig.~\ref{fig:InSb_temperature} (a) and (b). The data was fitted by the Drude model with fitted permittivity shown in Fig.~\ref{fig:InSb_temperature} (c) and (d). The fitting parameters are summarized in Fig.~\ref{fig:InSb_temperature} (e-g). 
At $T=70\,$K, the scattering rate is minimized as a consequence of the lower thermal free-carrier absorption and impurity scattering~\citep{ju2021tunable}. Therefore, the optimal temperature is set at $T_{\mathrm{opt}} = 70\,\mathrm{K}$ in the main text. 

\begin{figure}
    \centering
    \includegraphics[width = 3.25 in]{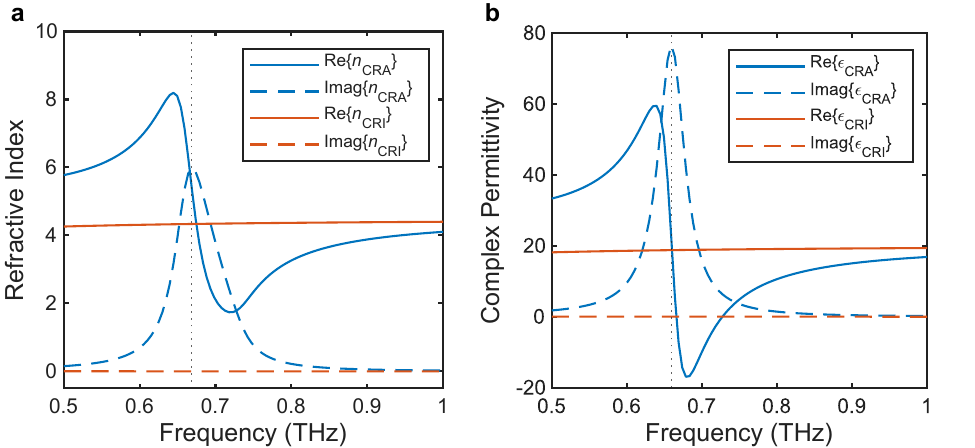}
    \caption{\textbf{Gyrotropic optical response of InSb in free space.} \textbf{a(b)},~The real and imaginary part of the complex refractive index (permittivity) of InSb at $T=70\,$K and $B=0.335\,$T, simulated from the Drude-Lorentz model. The simulation parameters are the same as shown in Fig.~\ref{fig:InSb_temperature} (e-g), and the effective mass is set at $m_{\mathrm{eff}}=0.014\,m_e$. Note the strong absorption for the CRA mode near 0.66\,THz and its absence for the CRI counterpart. }
    \label{fig:InSb_chiral}
\end{figure}

\section*{Chiral cavity transmittance at various temperatures and magnetic fields}

Here we present additional data on the THz response of the chiral PCC at different magnetic fields and temperatures. As in the main text, we consider two different configurations. 

\begin{figure}
    \centering
    \includegraphics[width=1\linewidth]{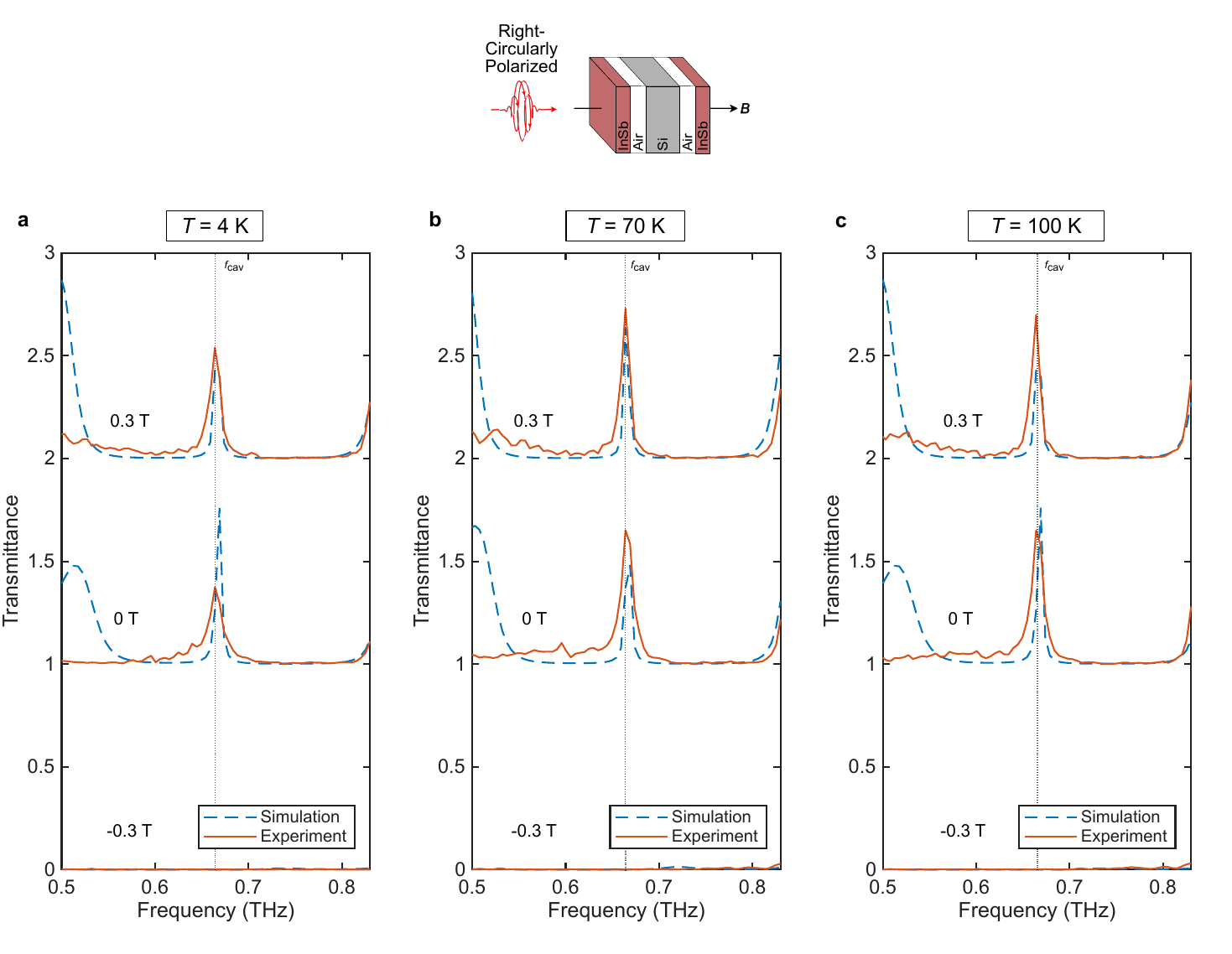}
    \caption{\textbf{Chiral photonic-crystal cavity transmittance with RCP incident light at different temperatures and magnetic fields. } \textbf{a-c},~The chiral PCC selectively transmits (absorbs) the RCP THz radiation at $B=0.3\,$T ($B=-0.3\,$T) and $T=4\,\mathrm{K}$, $70\,\mathrm{K}$, and $100\,\mathrm{K}$, consistent with the simulation (dashed blue line). }
    \label{fig:Cavity_RCP_T}
\end{figure}

In the first configuration, as shown in Fig.~\ref{fig:Cavity_RCP_T}, the RCP incident light is transmitted in a positive magnetic field $B=0.3\,$T and absorbed when $B=-0.3\,$T across a broad temperature range ($T=4\,\mathrm{K}$, $70\,\mathrm{K}$, and $100\,\mathrm{K}$). The transmittance agrees with the transfer-matrix-method simulation. 

\begin{figure}
    \centering
    \includegraphics[width=1\linewidth]{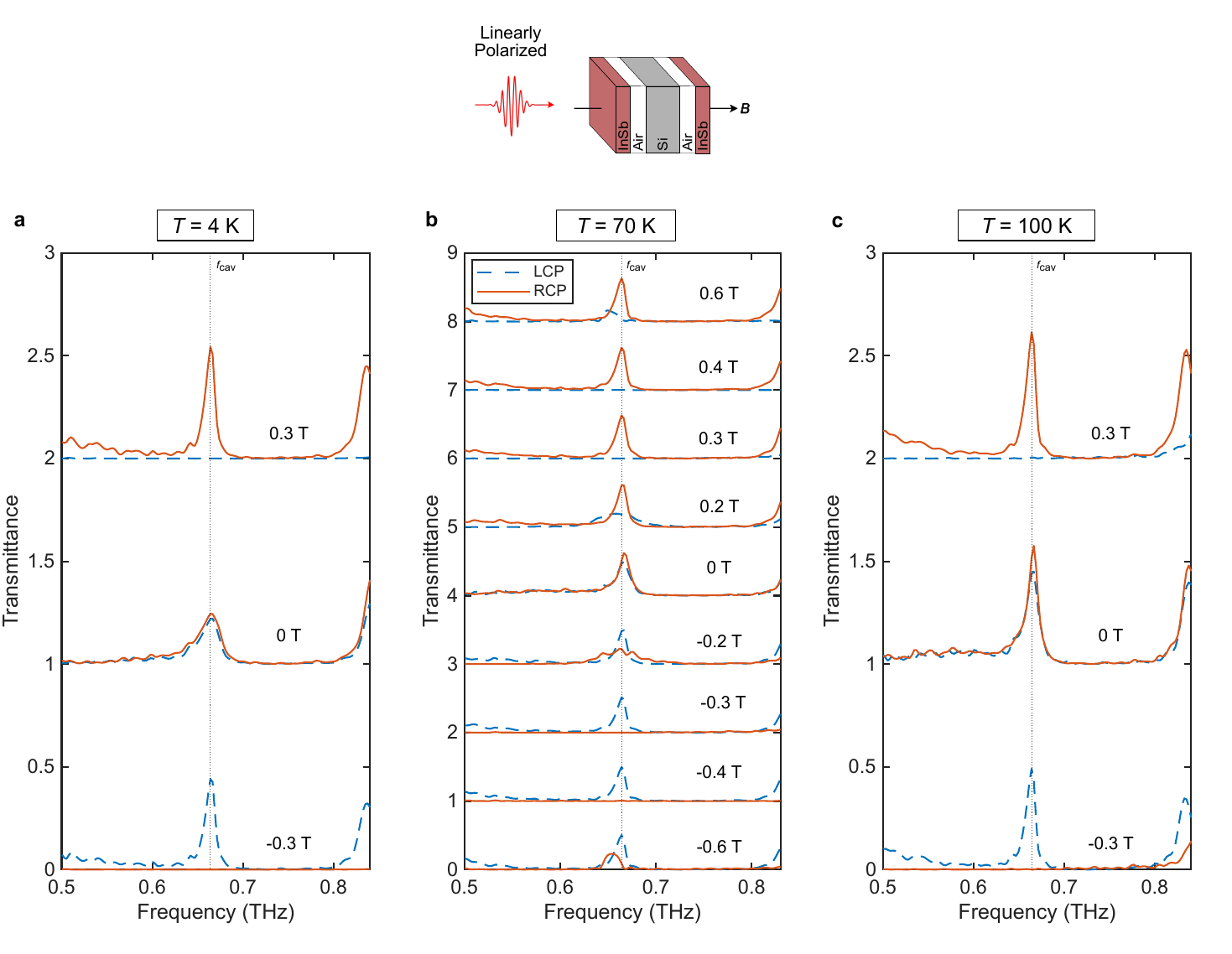}
    \caption{\textbf{Chiral photonic-crystal cavity transmittance with linear incident light. } \textbf{a-c},~The chiral PCC selectively transmits (absorbs) the RCP THz radiation at $B=0.3\,$T ($B=-0.3\,$T) and $T=4\,\mathrm{K}$, $70\,\mathrm{K}$, and $100\,\mathrm{K}$. For the LCP component (dashed blue line), the transmittance - magnetic field dependence is reversed.}
    \label{fig:Cavity_lin_T}
\end{figure}

In the second configuration, as shown in Fig.~\ref{fig:Cavity_lin_T}, the linearly-polarized incident THz light can be decomposed into the RCP and LCP components, and only the RCP (LCP) component transmits through the chiral PCC in a positive (negative) magnetic field $B=0.3\,$T ($B=-0.3\,$T). Similar to the first configuration, this polarization selectivity is robust at various temperatures. 

\section*{Stokes parameters and ellipticity calculations}

\begin{figure}
    \centering
    \includegraphics[width=1\linewidth]{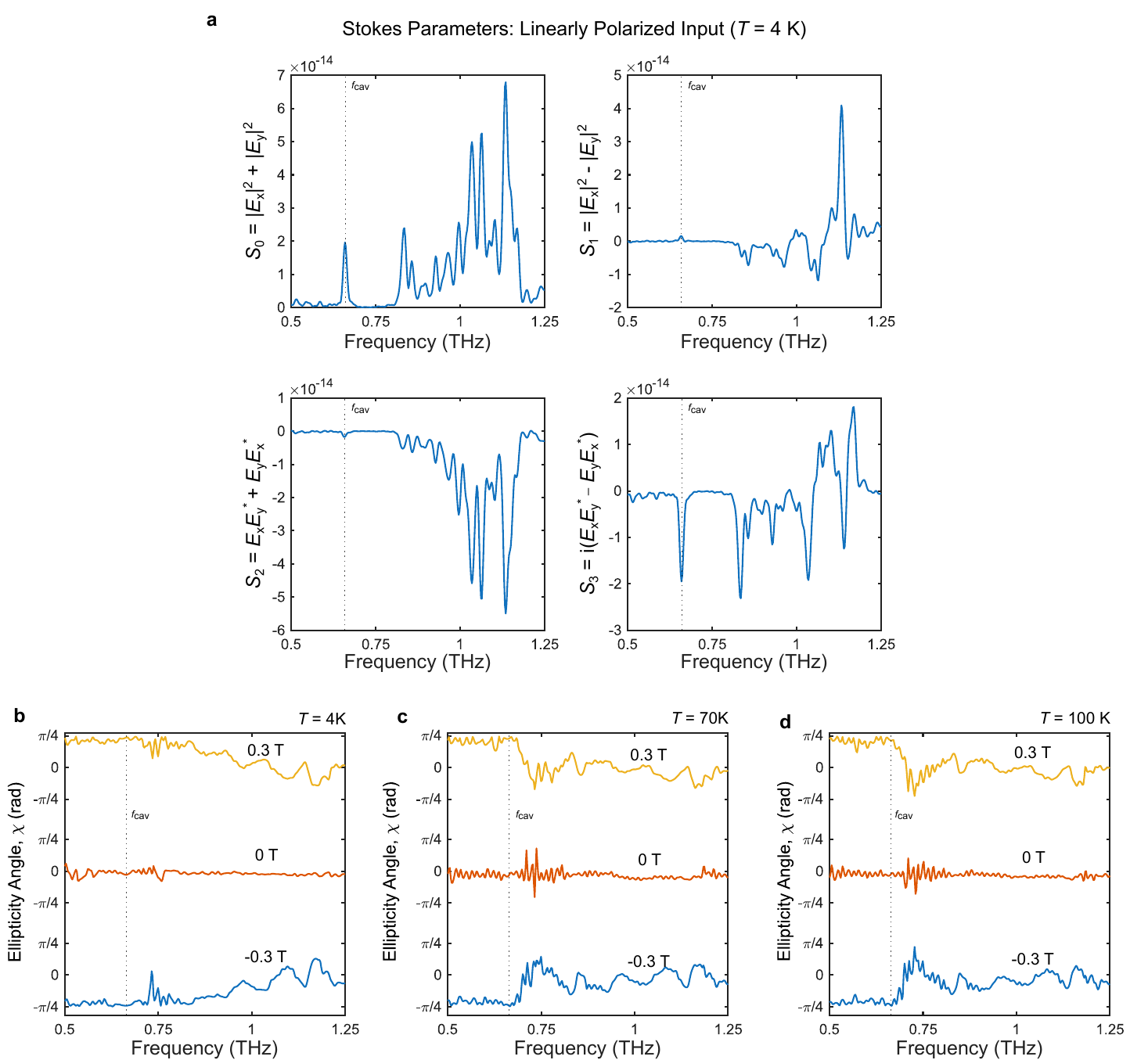}
    \caption{\textbf{Stokes parameters and ellipticity calculations.} \textbf{a},~The Stokes parameters as functions of frequency at \(B =0.3\,\)T and $T=4\,\mathrm{K}$. The vertical dashed line corresponds to the cavity mode near 0.66\,THz. \textbf{b-d},~The ellipticity obtained from Stokes parameters at $T=4\,\mathrm{K}$, $70\,\mathrm{K}$, and $100\,\mathrm{K}$. The vertical dashed line corresponds to the cavity mode near 0.66\,THz.}
    \label{fig:Stokes}
\end{figure}

As shown in Fig.~\ref{fig:Stokes}, the Stokes parameters are calculated based on the transmitted complex linearly-polarized electric field components obtained from the THz-TDS measurements with the formulae in the Methods section of the main text. We can then calculate the ellipticity from the Stokes parameters. We show that at various temperatures the ellipticity at the cavity mode frequency is close to the theoretical maximum value of $\pm \pi / 4$, demonstrating high circular dichroism across a wide temperature range. 

\section*{\(Q-\)factor calculation}
\begin{figure}
    \centering
    \includegraphics[width=1.75in]{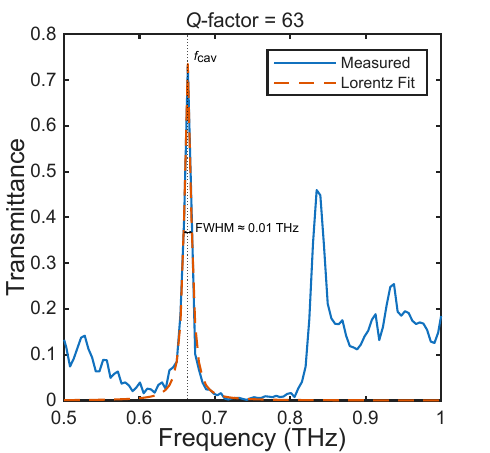}
    \caption{\textbf{Cavity \(Q\)-factor calculation.} Lorentzian fit of the cavity mode resonance at \(T = 70\,\)K and \(B = 0.3\,\)T under RCP incidence. The extracted full width at half maximum (\(\approx 0.1\,\){THz}) corresponds to a quality factor \(Q \approx 63\).}
    \label{fig:Q-factor}
\end{figure}

As shown in Fig. \ref{fig:Q-factor}, the cavity mode transmittance is fitted by a Lorentzian of the form:
\begin{equation}
    L(f) \;=\; \frac{A}{\pi} \frac{\gamma}{(f - f_{\rm cav})^2 + \gamma^2}.
\end{equation}
Here, \(A\) is a scaling factor, \(2\gamma\) is the full-width at half maximum of the cavity mode (FWHM), and \(f_{\rm cav}\) is the cavity resonance frequency. From this relation, the cavity quality factor is obtained as
\begin{equation*}
    Q = \frac{f_{\rm cav}}{2\gamma}.
\end{equation*}

\begin{figure}
    \centering
    \includegraphics[width=3.25in]{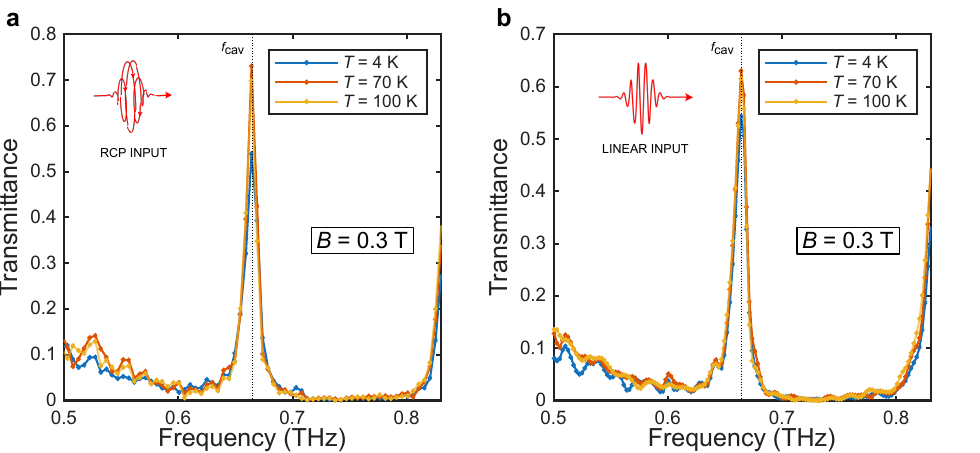}
    \caption{\textbf{Cavity mode transmittance peak as a function of \(T\)}. The chiral cavity confines the cavity mode robustly at various temperatures. The maximum transmittance of cavity mode appears \(\approx 70\,\)K, confirming the optimum temperature trend in \ref{fig:InSb_temperature} e-g. }
    \label{fig:PeakvsT}
\end{figure}

\end{document}